\documentclass[twoside,twocolumn,9pt]{article}
\usepackage{extsizes}
\usepackage[super,sort&compress,comma]{natbib}
\usepackage[version=3]{mhchem}
\usepackage[left=1.5cm, right=1.5cm, top=1.785cm, bottom=2.0cm]{geometry}
\usepackage{balance}
\usepackage{times,mathptmx}
\usepackage{sectsty}
\usepackage{graphicx}
\usepackage{lastpage}
\usepackage[format=plain,justification=justified,singlelinecheck=false,font={stretch=1.125,small,sf},labelfont=bf,labelsep=space]{caption}
\usepackage{float}
\usepackage{fancyhdr}
\usepackage{fnpos}
\usepackage[english]{babel}
\addto{\captionsenglish}{%
  }
\usepackage{array}
\usepackage{droidsans}
\usepackage{charter}
\usepackage[T1]{fontenc}
\usepackage[usenames,dvipsnames]{xcolor}
\usepackage{setspace}
\usepackage[compact]{titlesec}
\usepackage{hyperref}
\usepackage{bm}
\usepackage{latexsym}
\usepackage{amsfonts}
\usepackage{graphicx}
\usepackage{subfigure}
\usepackage{epsfig}
\usepackage{setspace}
 \usepackage{natbib}
\usepackage{dcolumn}

\usepackage{epstopdf}
\definecolor{cream}{RGB}{222,217,201}

\begin{document}

\pagestyle{fancy}
\thispagestyle{plain}

\fancypagestyle{plain}{

\fancyhead[C]{\includegraphics[width=18.5cm]{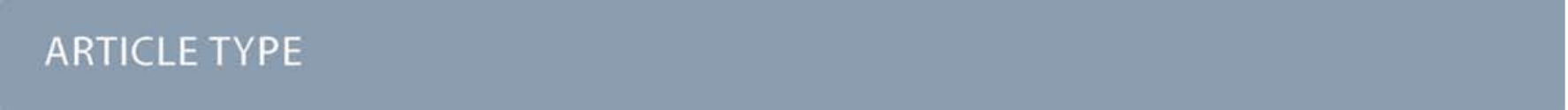}}
\fancyhead[L]{\hspace{0cm}\vspace{1.5cm}\includegraphics[height=30pt]{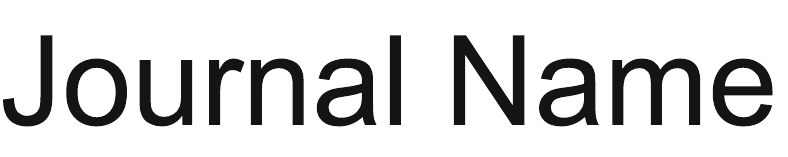}}
\fancyhead[R]{\hspace{0cm}\vspace{1.7cm}\includegraphics[height=55pt]{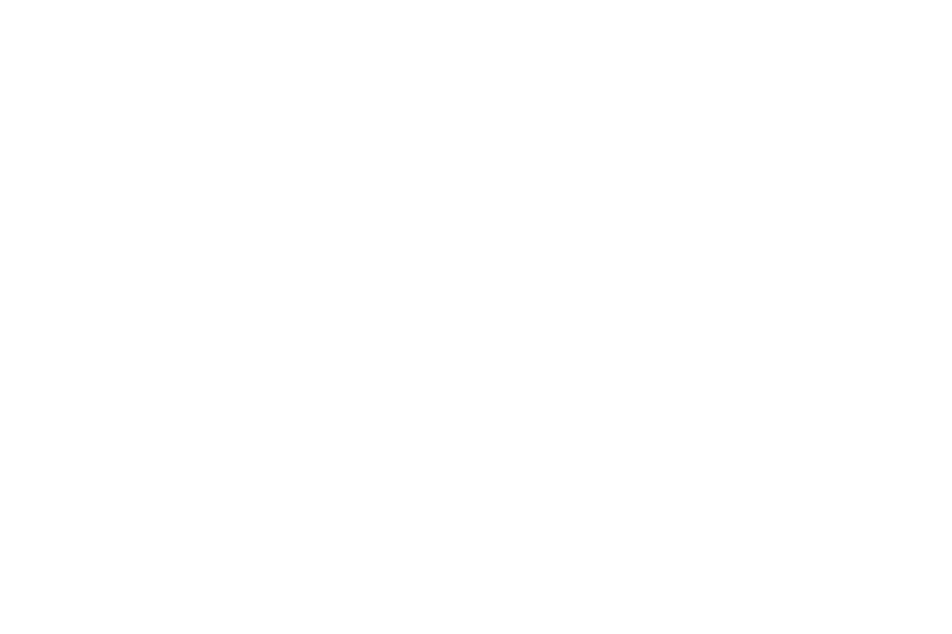}}
\renewcommand{\headrulewidth}{0pt}
}

\makeFNbottom
\makeatletter
\renewcommand\LARGE{\@setfontsize\LARGE{15pt}{17}}
\renewcommand\Large{\@setfontsize\Large{12pt}{14}}
\renewcommand\large{\@setfontsize\large{10pt}{12}}
\renewcommand\footnotesize{\@setfontsize\footnotesize{7pt}{10}}
\makeatother

\renewcommand{\thefootnote}{\fnsymbol{footnote}}
\renewcommand\footnoterule{\vspace*{1pt}%
\color{cream}\hrule width 3.5in height 0.4pt \color{black}\vspace*{5pt}}
\setcounter{secnumdepth}{5}

\makeatletter
\renewcommand\@biblabel[1]{#1}
\renewcommand\@makefntext[1]%
{\noindent\makebox[0pt][r]{\@thefnmark\,}#1}
\makeatother
\renewcommand{\figurename}{\small{Fig.}~}
\sectionfont{\sffamily\Large}
\subsectionfont{\normalsize}
\subsubsectionfont{\bf}
\setstretch{1.125} 
\setlength{\skip\footins}{0.8cm}
\setlength{\footnotesep}{0.25cm}
\setlength{\jot}{10pt}
\titlespacing*{\section}{0pt}{4pt}{4pt}
\titlespacing*{\subsection}{0pt}{15pt}{1pt}

\fancyfoot{}
\fancyfoot[LO,RE]{\vspace{-7.1pt}\includegraphics[height=9pt]{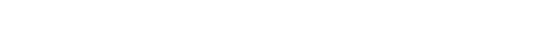}}
\fancyfoot[CO]{\vspace{-7.1pt}\hspace{13.2cm}\includegraphics{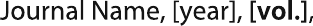}}
\fancyfoot[CE]{\vspace{-7.2pt}\hspace{-14.2cm}\includegraphics{head_foot/RF}}
\fancyfoot[RO]{\footnotesize{\sffamily{1--\pageref{LastPage} ~\textbar  \hspace{2pt}\thepage}}}
\fancyfoot[LE]{\footnotesize{\sffamily{\thepage~\textbar\hspace{3.45cm} 1--\pageref{LastPage}}}}
\fancyhead{}
\renewcommand{\headrulewidth}{0pt}
\renewcommand{\footrulewidth}{0pt}
\setlength{\arrayrulewidth}{1pt}
\setlength{\columnsep}{6.5mm}
\setlength\bibsep{1pt}

\makeatletter
\newlength{\figrulesep}
\setlength{\figrulesep}{0.5\textfloatsep}

\newcommand{\topfigrule}{\vspace*{-1pt}%
\noindent{\color{cream}\rule[-\figrulesep]{\columnwidth}{1.5pt}} }

\newcommand{\botfigrule}{\vspace*{-2pt}%
\noindent{\color{cream}\rule[\figrulesep]{\columnwidth}{1.5pt}} }

\newcommand{\dblfigrule}{\vspace*{-1pt}%
\noindent{\color{cream}\rule[-\figrulesep]{\textwidth}{1.5pt}} }

\makeatother

\twocolumn[
  \begin{@twocolumnfalse}
\vspace{3cm}
\sffamily
\begin{tabular}{m{4.5cm} p{13.5cm} }

\includegraphics{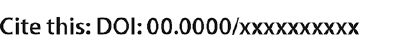} & \noindent\LARGE{\textbf{Effect of the interaction strength and anisotropy on the diffusio-phoresis of spherical colloids}} \\
\vspace{0.3cm} & \vspace{0.3cm} \\

 & \noindent\large
 {Jiachen Wei,\textit{$^{a,b}$}
 Sim\'on Ram\'irez-Hinestrosa,\textit{$^{b}$}
 Jure Dobnikar, \textit{$^{b,c,d\ast}$}
 and Daan Frenkel\textit{$^{b\dag}$}}
 \\

\includegraphics{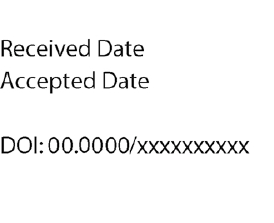} & \noindent\normalsize{Gradients in temperature, concentration or electrostatic potential cannot exert forces on a bulk fluid; they can, however, exert forces on a fluid in a microscopic boundary layer surrounding a (nano)colloidal solute, resulting in so-called phoretic flow.
Here we present a simulation study of phoretic flow around a spherical colloid held fixed in a concentration gradient.
We show that the resulting flow velocity depends non-monotonically on the strength of the colloid-fluid interaction.
The reason for this non-monotonic dependence is that solute particles are effectively trapped in a shell around the colloid and cannot contribute to diffusio-phoresis.
We also observe that the flow depends sensitively on the anisotropy of solute-colloid interaction.  } \\

\end{tabular}

 \end{@twocolumnfalse} \vspace{0.6cm}
]

\renewcommand*\rmdefault{bch}\normalfont\upshape
\rmfamily
\section*{}
\vspace{-1cm}

\footnotetext{\textit{$^{a}$~Institute of Mechanics, Chinese Academy of Sciences, Beijing 100190, China }}
\footnotetext{\textit{$^{b}$~Department of Chemistry, University of Cambridge, CB21EW Cambridge, UK}}
\footnotetext{\textit{$^{c}$~Institute of Physics, Chinese Academy of Sciences, Beijing 100190, China}}
\footnotetext{\textit{$^{d}$~Songshan Lake Materials Laboratory, Dongguan 523808, China}}
\footnotetext{\textit{$^{\ast}$~jd489@cam.ac.uk}}
\footnotetext{\textit{$^{\dag}$~df246@cam.ac.uk}}


\section{\label{sec:Intro}Introduction}

The term ``phoresis'' covers  a class of transport phenomena that are generated by thermodynamic gradients.
Phoretic transport can be induced by gradients in chemical potential~\cite{Eijkel10, Anderson06, Annunziata12, Ault17, Deseigne14, Khair13, Palacci10, Paustian15, Prieve18, Sear17, Shi16, Shin17, Velegol16, Yang14}, temperature~\cite{Wurger10, Braibanti08, Burelbach17, Burelbach17-2, Fu17, Maeda11, Piazza08, Piazza08-2, Tsuji17, Vigolo10, Vigolo17, Wienken10, Wolff16} or electrostatic potential~\cite{Molotilin16, Semenov13, Stout17, Tanaka02, Tanaka03}.
Electrophoresis is commonly used to separate bio-molecules, but other phoretic phenomena could, in principle, be used for the same purpose.
In this paper, we consider the challenges involved in developing separation techniques based on diffusio-phoresis.

The conventional theoretical description of phoretic transport combines a local thermodynamics description of the fluid around a colloidal particle with Stokesian hydrodynamics~\cite{Anderson81, Anderson06, Khair13, Yang14, Yang17, Chen17}.
Whilst such a continuum description is usually adequate  for electrophoresis, it is less applicable in the case of diffusio- and thermo-phoresis, where the characteristic length-scales are often too small to justify local hydrodynamics/thermodynamics.
In particular, the continuum picture fails to account for ordering on an atomistic scale
near an interface~\cite{Huang16, Liu18}, even though its predictions are qualitatively similar to those obtained by molecular simulation~\cite{Sharifi-Mood13}.
As experiments move increasingly to the nano-scale~\cite{Abecassis08, Banerjee16, Saar17}, and as direct simulations are becoming feasible~\cite{Yang16}, we are now in a position to explore the diffusion-phoretic transport of uncharged (nano)colloidal particles.

The most straightforward method to simulate phoresis is to impose an explicit gradient in the concentration or temperature~\cite{Sharifi-Mood13, Liu17, Liu18, Fu17}.
This approach is less suited for systems with periodic boundary conditions.
However, as shown in recent papers, the gradient can also be imposed as an external force  acting on the chemical identity (in the case of diffusio-phoresis) or the excess enthalpy (in the case of thermophoresis) of individual particles~\cite{Yoshida17, Ganti17}.

In the present paper we investigate the dependence of the diffusio-phoretic mobility on the strength and  anisotropy of the solute-colloid interaction.
In addition, we consider the effect of different hydrodynamic boundary conditions.
In the next section we introduce the model and the simulation methods and then proceed to analyze the results of the simulations.
\begin{figure}[h!]
   \begin{center}
      \includegraphics*[width=0.46\textwidth]{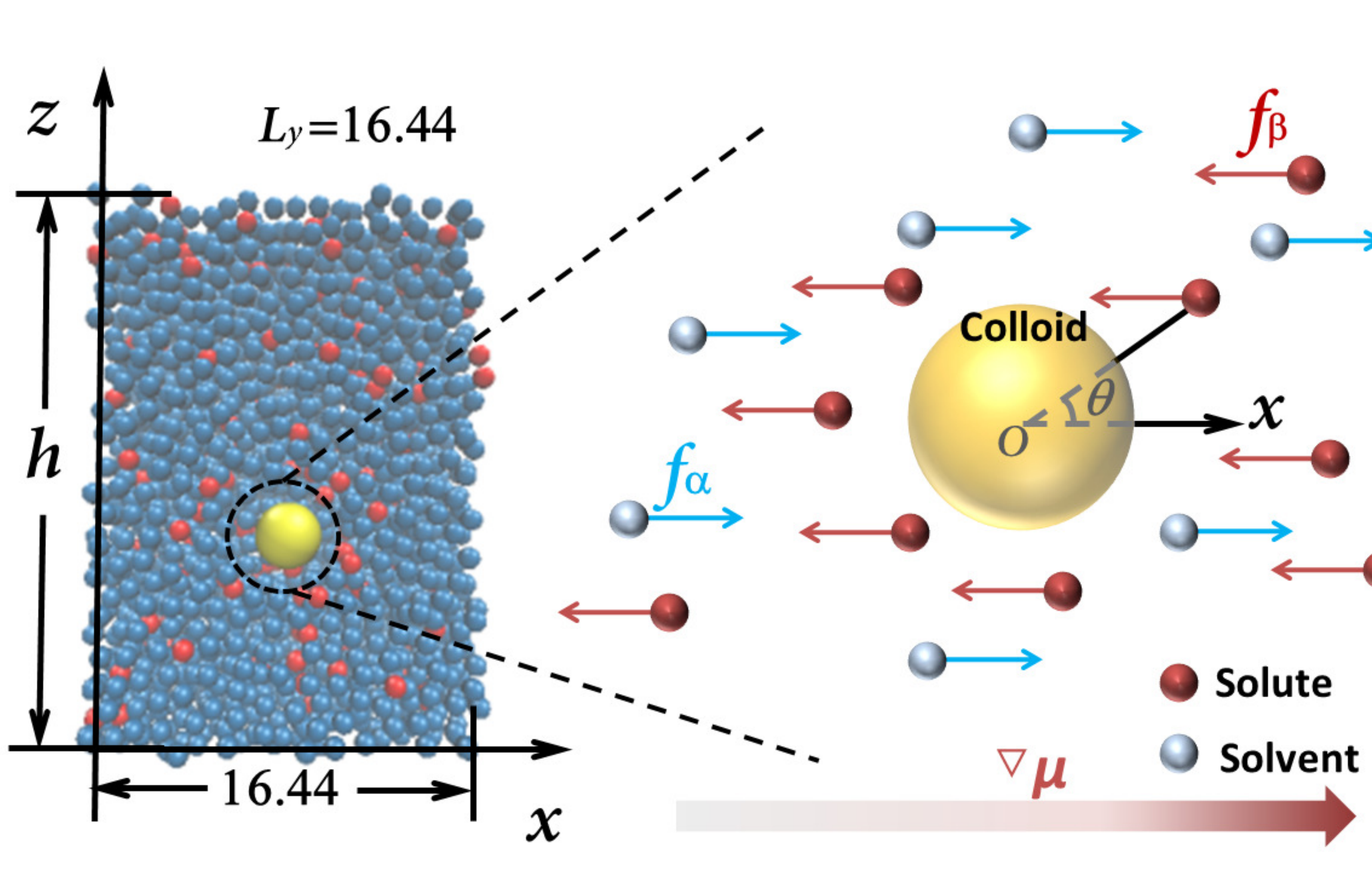}
      \caption{The left figure shows a colloidal particle embedded in a mixture of solvent (blue) and solute (red) particles. The right-hand figure shows the gradient in the chemical potential of solvent and solute particles,  represented by ``color'' forces acting on these particles. The strength of these forces is such that the average force on a fluid element in the bulk of the liquid vanishes, because the bulk pressure is constant throughout the system.}
      \label{fig:model}
   \end{center}
\end{figure}

\section{\label{sec:Method}Simulation method}

\subsection{\label{sec:Met2Model}Model}

We performed Molecular Dynamics (MD) simulations  of  diffusio-phoresis in a system with a single colloidal particle ($c$) immersed in a fluid comprising solvent ($\alpha$) and solute ($\beta$) particles.
All fluid particles were assumed to be of the same diameter $\sigma_s$, which was used as a unit of length.
The diameter of the colloidal particle was taken to be $\sigma_c=3$.
The system dimensions  were $16.44 \times 16.44 \times h$, where the box height $h$ was varied to keep the pressure constant.
We applied periodic boundary conditions and measured the velocity of the fluid flow, whilst keeping the colloidal particle fixed at the origin.

We assumed interaction potentials of the form
\begin{equation}\label{eqn:Uint}
U_{ij}(\mathbf{r}_i,\mathbf{r}_i) = U^{SR}_{ij} (r_{ij}) + \lambda_{ij} U^{A}_{ij}(r_{ij},\theta_{ij})\;,
\end{equation}
where $U^{SR}(r)$ denotes an isotropic short range potential $U^{SR}(r)$, while $U^{A}(r,\theta)$ denotes a possible anisotropic interaction between colloid and solutes.
We assume that the fluid particles interact isotropically among themselves and that the only interaction with angular dependence is that between the colloid and the solute particles.
Therefore, the only non-vanishing pre-factor controlling the strength of the anisotropy is $\lambda_{\beta c}\equiv\lambda_B$.
The choice of the short-ranged  isotropic interaction potential was dictated by computational convenience. We start from a recently introduced generic, short-ranged, attractive pair potential~\cite{Wang2019}:
\[
u(r_{ij})\equiv a_{ij}\left[\left(\frac{\sigma_{ij}}{r_{ij}}\right)^2-1\right]\left[\left(\frac{r_{cut}}{r_{ij}}\right)^2-1\right]^{2}\;.
\]
where $a_{ij}$ is chosen such that the minimum value of $u(r_{ij})$ equals -1:
\begin{equation}\label{eqn:alpha}
a _{ij}\equiv 2\left(\frac{r_{cut}}{\sigma_{ij}}\right)^2\left[\frac{3}{2\left(r_{cut}{}^2/\sigma_{ij}{}^2-1\right)}\right]^{3}.
\end{equation}
From this pair potential, we construct a short-ranged potential $U^{\scriptscriptstyle SR}_{ij}$ that has an adjustable well depth $\epsilon_{ij}$:
\begin{equation}\label{eqn:ESR}
 U^{\scriptscriptstyle SR}_{ij} = \left\{ \begin{array}{ll}
       u(r_{ij})+1-\epsilon_{ij} & \textrm{for $r_{ij}<r^m_{ij}$}, \\
      \epsilon_{ij} u(r_{ij}) & \textrm{for $r^m_{ij} \leq r_{ij}<r_{cut}$}, \\
      0 & \textrm{for $r_{ij}\geq r_{cut}$}\;.
  \end{array} \right.
\end{equation}
Indices denote the type of the particles: $i,j \in \{c,\alpha,\beta\}$, $\sigma_{ij}$ is the distance where the potential crosses zero, $r_{cut}$ denotes the cutoff radius (we choose $r_{cut}=2$ for fluid--fluid, and $r_{cut}=5$ for fluid--colloid interactions), $r_{ij}^m$ the location of the potential minimum, and $\epsilon_{ij}$ controls the depth of the potential well.

In what follows, the strength of the interaction between all fluid species as our unit of energy: $\epsilon_{\alpha\alpha}=\epsilon_{\alpha\beta}=\epsilon_{\beta\beta}\equiv 1$.
Furthermore, we fix the solvent-colloid interaction strengths, $\epsilon_{\alpha c}=1$, and vary the solute-colloid interaction $\epsilon_{\beta c}\equiv \epsilon_B$).
The overview of the interaction parameters is shown in the Table~\ref{table2}.
\begin{table}[h!]
\centering
\begin{tabular}{|l|l|l|l|}
\hline
$i j$ & $\epsilon$ & $\sigma$ & $\lambda$ \\
\hline
 solvent-solvent  &  $\epsilon_{\alpha\alpha}=1.0$  &  $\sigma_{\alpha\alpha}=1.0$ & $\lambda_{\alpha\alpha}=0$ \\
 solvent-solute & $\epsilon_{\alpha\beta}=1.0$  &  $\sigma_{\alpha\beta}=1.0$  & $\lambda_{\alpha\beta}=0$\\
 solute-solute & $\epsilon_{\beta\beta}=1.0$  &  $\sigma_{\beta\beta}=1.0$  & $\lambda_{\beta\beta}=0$\\
\hline
solvent-colloid &  $\epsilon_{\alpha c}=0$  &  $\sigma_{\alpha c}=3.0$ & $\lambda_{\alpha c}=0$ \\
solute-colloid & $\epsilon_{\beta c}\equiv\epsilon_{B}$  &  $\sigma_{\beta c}=3.0$ & $\lambda_{\beta c}\equiv\lambda_B$ \\
\hline
\end{tabular}
\caption{This table lists the interaction parameters used in the simulations.
The solvent-solvent, solvent-solute and solute-solute interactions are isotropic ($\lambda=0$) and have the same values of $\sigma$ and $\epsilon$, which are always kept constant.
The solvent-colloid interaction is purely repulsive  ($\epsilon_{\alpha c}=0$)  and isotropic ($\lambda=0$).
For the solute-colloid interaction, both the strength of the attraction ($\epsilon_{\beta c}$) and the strength of the anisotropic interaction ($l=1$ or $l=2$), are varied. \label{table2}}
\end{table}

The minimum of the interaction potential is located at:
\begin{equation}\label{eqn:rm}
 r_{ij}^m=r_{cut}\sqrt{\frac{3}{1+2r_{cut}{}^2/\sigma_{ij}{}^2}}.
\end{equation}
An advantage of this model  potential is that the potential and its first derivative vanish at $r_{cut}$~\cite{Wang2019}. However, for the solute-colloid interaction, the force is discontinuous  for $0<\epsilon_{\beta c}<1$, although the potential is continuous.

When studying the effect of anisotropic colloid-solute interactions, we decompose these interactions in terms of Legendre polynomials $P_l(\cos\theta)$:
\begin{equation}\label{eqn:Aniso}
 U^{\scriptscriptstyle A}_{\beta c} = \left\{ \begin{array}{ll}
      -P_l(\cos\theta) & \textrm{for $r_{\beta c}<r^m_{\beta c}$}, \\
      P_l(\cos\theta) u(r_{\beta c}) & \textrm{for $r^m_{\beta c} \leq r_{\beta c}<r_{cut}$}, \\
      0 & \textrm{for $r_{\beta c}\geq r_{cut}$},
  \end{array} \right.
\end{equation}
where $\theta$ denotes the angle between symmetry axis of the colloidal particle and the vector joining the centers of mass of the colloid and a solute particle (see Fig.~\ref{fig:model}). The advantage of using a Legendre-polynomial decomposition is that, in the limit of weak anisotropic interactions, the integrated excess density vanishes for all $l \ge 1$.

As shown in Fig.~\ref{fig:model}, we represent the effect of a concentration (or chemical-potential) gradient~\cite{Yoshida17, Liu17} by applying opposing external ``color" forces $f_i$ on the solvent and solute particles:
\begin{equation}\label{eqn:force}
f_i=\left(\frac{-\partial \mu _i^{\text{bulk}} }{\partial \rho _i}\right)_P\cdot \nabla \rho _i,
\end{equation}
where $\mu^{\text{bulk}}$ is the chemical potential and $\rho$ the number density.
The reason for carrying out a simulation with color forces, rather than the corresponding explicit concentration gradients, is twofold. First of all, as in the case of simulations of systems in electrical fields, it is often better to impose a constant field that is compatible with the periodic boundary conditions, than to impose a periodic charge density that would locally lead to the same field.

The second reason is specific for phoretic transport: in principle, we should carry out these simulations under conditions where the phoretic flow velocity is so small that the concentration profile is not perturbed by the flow (the limit of vanishing Peclet number). With explicit concentration gradients, the relevant Peclet number is $vL/D$, where $v$ is te average phoretic flow velocity, $L$ is the system size and $D$ is the diffusion coefficient of the solutes/solvent molecules. However, with imposed color forces, the relevant Peclet number is $v\sigma/D$, where $\sigma$ is the colloidal diameter. As, typically, $\sigma\ll L$, the Peclet number for simulations with color forces is much smaller than the corresponding number in the presence of explicit concentration gradients. As shown in the SI, simulations with an explicit concentration gradient~\cite{Liu17}, in a $36.17 \times 16.44 \times h$ box with the position of the colloidal particle fixed at the origin, lead to a strongly non-linear (in fact: exponential) concentration profile at flow velocities where no such problem occurs if we impose color forces.

\subsection{\label{sec:Met2Sim}Simulation Details}
All simulations were performed at constant $N$, $P$ and $T$, with the number of particles fixed at $N=4836$,  and the temperature at $T=0.845$.
The length of the simulation time step was $dt=0.001$, and a Nos\'{e}-Hoover thermostat with a time constant of $100\,dt$ was used to control the temperature.
Here and in what follows, we use reduced units, based on the $\epsilon$, $\sigma$ and $m$ (mass) of the solvent particles.

All initial configurations were prepared by placing  $N_\alpha=4835$ solvent particles on a low-density FCC crystal lattice.
$N_\beta=\rho_\beta V$ randomly chosen solvent particles were then replaced with solute particles, where $V$ is the volume of the simulation box.
The system was then compressed in $z$ direction and quenched to the desired pressure ($P=0.012$).
The identities of solvent and solute particles in the bulk (i.e. far away from the colloidal particle) were allowed to interchange every $500$ simulation steps to keep the solute density constant at $\rho_\beta = 0.381$ for at least $4\times10^7$ simulation steps.
Color forces $f_i$ along the $x$-direction were then applied to the solvent and solute particles.
Subsequently, a run of at least $2\times10^8$ steps was performed to obtain the flow rate and density profile,  which were collected by averaging over $10^5$ output configurations separated by $10^3$ simulation steps.
The applied color forces correspond to a solute concentration gradient of  $|\nabla\rho_\beta|=0.04$.

\begin{figure}[h!]
   \begin{center}
      \includegraphics*[width=0.38\textwidth]{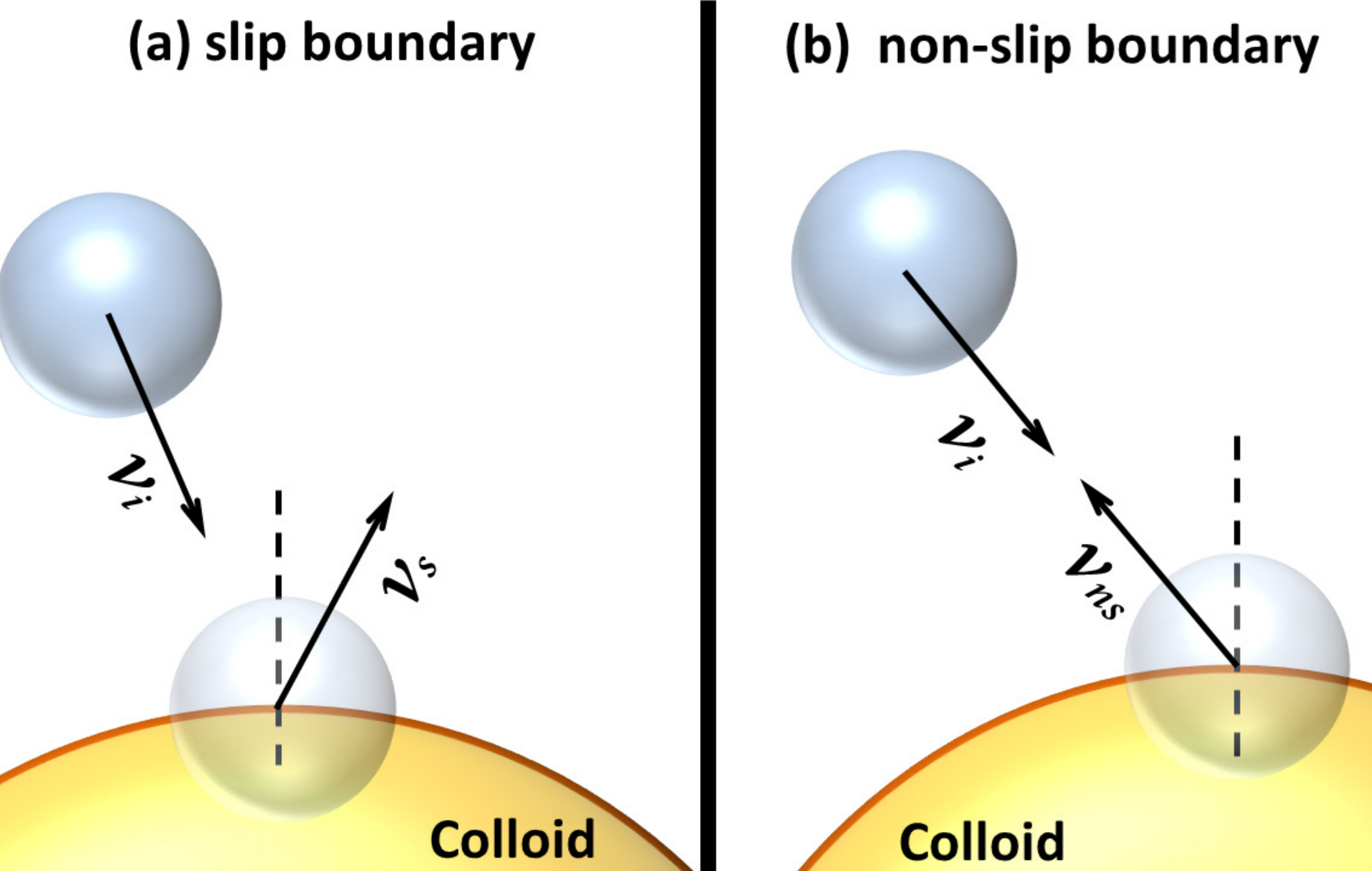}
      \caption{Schematic illustration of the implementation of slip (a) and non-slip (b) boundary conditions.}
      \label{fig:boundary}
   \end{center}
\end{figure}

Integrating the MD equations of motion with the Verlet algorithm, which  conserves the tangential velocity of a solvent/solute particle when interacting with the colloid.
As a result, this situation corresponds to a ``slip'' boundary condition (Fig.~\ref{fig:boundary}(a)). To implement non-slip boundary conditions, we reverse the velocity of the solvent/particle at the distance $r^{ns}$ (Fig.~\ref{fig:boundary}(b)).
In our simulations, we chose $r^{ns}=\sigma_{ic}$, where $i=(\alpha,\beta)$ refers to both solvent and solute particles.
Note, however, that the definition of non-slip boundary conditions is not unambiguous for systems with a  continuous colloid-fluid interaction.

\section{\label{sec:ReDis}Result and discussion}
\subsection{\label{sec:Res1}Isotropic interaction}
\begin{figure}[h!]
   \begin{center}
      \includegraphics*[width=0.5\textwidth]{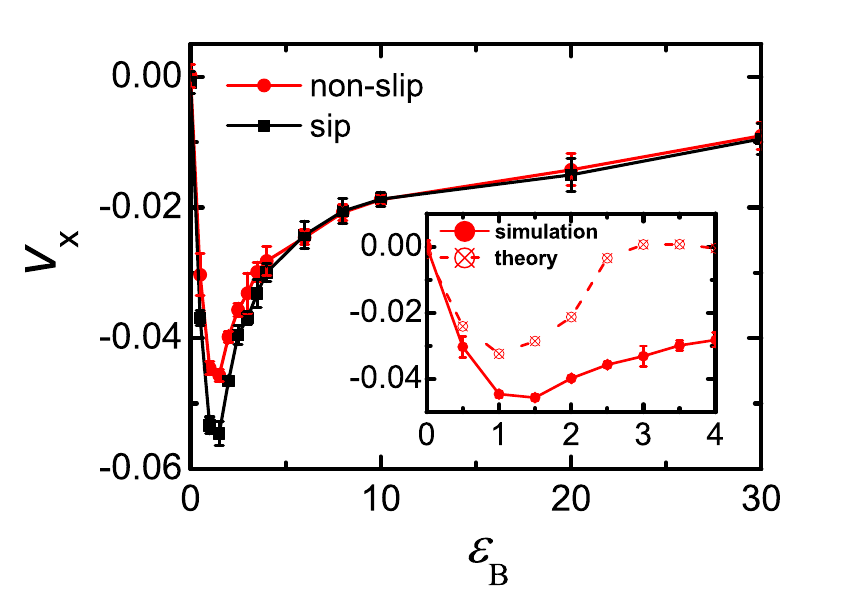}
      \caption{Fluid velocity  $v_x$ versus solute-colloid (isotropic) interaction strength   $\epsilon_B$ for slip (red; circles) and non-slip (black; squares) boundaries. The inset shows a comparison of the non-slip data in the regime $\epsilon_B<4.0$ (solid) to the Derjaguin--Anderson theory (Eq.\ref{eqn:vcorrect}, dashed). }
      \label{fig:nons}
   \end{center}
\end{figure}

Fig.~\ref{fig:nons} shows the diffusio-phoretic flow past a colloid that interacts with the solvent and solute particles with isotropic ($\lambda_B=0$) potential specified in the previous section.
For $\epsilon_B < 1.5$, the magnitude of the velocity in the direction of phoretic motion obtained from simulations, $v_x$, increases with $\epsilon_B$.
This is because the number of adsorbed solute particles, and hence total color force exerted on a fluid element close to the colloidal surface, increases with $\epsilon_B$.
Beyond $\epsilon_B\approx 1.5$ $v_x$ decreases with $\epsilon_B$ and reaches a plateau at $v_x\sim-0.02$.
Stronger attraction between the colloid and solute implies a larger excess of solute near the colloid.
However, as the colloid-solute attraction becomes stronger,  the solute particles become less mobile and cannot participate in diffusio-phoretic flow (similar behaviour was observed for polymer diffusio-phoresis see~\cite{Ramirez2019}).

\begin{figure}[h!]
   \begin{center}
      \includegraphics*[width=0.5\textwidth]{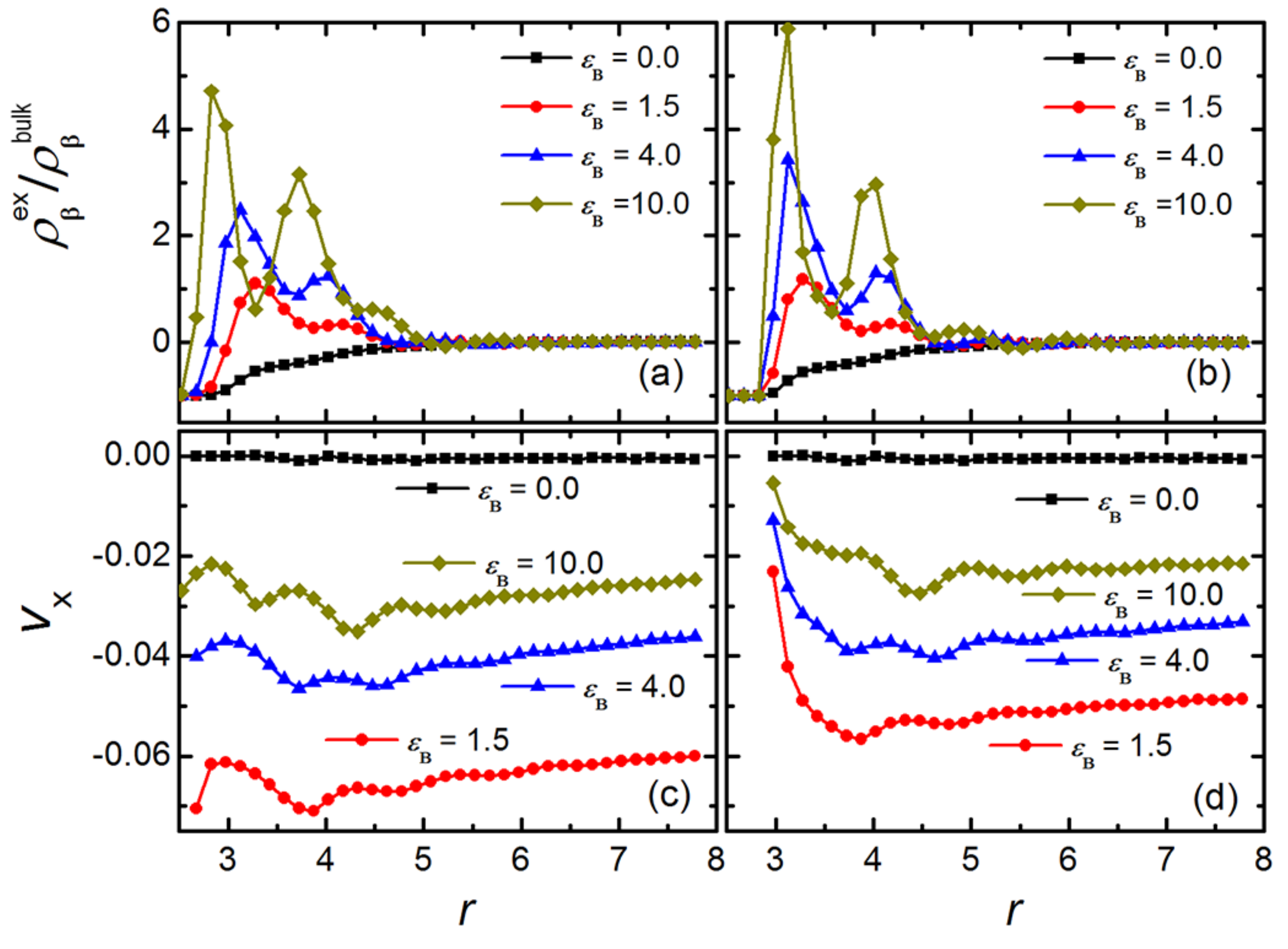}
      \caption{The radial distribution of the reduced excess density of solutes $\rho^{ex}_\beta/\rho^{bulk}_\beta$ with (a) slip and (b) non-slip boundaries, and the radial function of fluid velocity $v_x$ with (c) slip and (d) non-slip boundaries.}
      \label{fig:P0allnew}
   \end{center}
\end{figure}

We  compare our simulation results with the Derjaguin--Anderson theory for diffusio-phoresis~\cite{Anderson81}. The theory predicts the phoretic flow around a spherical colloid with the hydrodynamic radius  $R_h$ at constant fluid viscosity $\eta$:
\begin{equation}\label{eqn:vcorrect}
v_x=v_x^0\left(1-\frac{K+H}{R_h}\right),
\end{equation}
where  $v_x^0={\textstyle \frac{k_BT}{\eta }} \int _{R_h}^{\infty }r dr \left [\rho _{\beta }^{ex}(r)\nabla \rho _{\beta }/\rho _{\beta } +\rho _{\alpha }^{ex}(r) \nabla \rho _{\alpha }/\rho _{\alpha } \right ]$  is the flow velocity in the Derjaguin limit assuming a flat surface, with $\rho_\beta^{ex}$ and $\rho_\alpha^{ex}$ denoting the excess density of solutes and solvents, respectively. The Anderson curvature corrections are introduced via the functions $K=\int _{R_h}^{\infty }  \omega(r) dr$ and $H=\int _{R_h}^{\infty }  \omega(r) \frac{r^2}{2} dr / \int _{R_h}^{\infty }  \omega(r) r dr$ with $\omega(r) \equiv \rho _{\beta }^{ex}(r) / \rho _{\beta } - \rho _{\alpha }^{ex}(r) / \rho _{\alpha}$. In order to obtain the theoretical prediction for the phoretic flow $v_x$ as a function of the interaction strength $\epsilon_B$, we have first evaluated the hydrodynamic radius of the colloid by measuring its mean-squared displacement and used this value in Eq.~\ref{eqn:vcorrect}. The comparison between the Derjaguin--Anderson prediction and the simulation results with non--slip boundaries is shown in the inset of Fig.~\ref{fig:nons}. Qualitatively, the behaviour is similar, although there is an important difference, {\sl i.e.}, the theoretical prediction for the flow goes to zero at moderate $\epsilon_B$, while the simulation data seem to converge to a finite plateau. The Derjaguin--Anderson theory ignores the variation of the local viscosity and assumes a sharp profile with no flow within a well-defined hydrodynamic radius of the colloid, $R_h$. As we demonstrate later in Fig.\ref{fig:P0allnew}, the actual concentration and flow profiles are not sharp at all, which is why the theory does not  account quantitatively for the simulation data.

Fig.~\ref{fig:nons} shows that the non-monotonic relation between $v_x$ and $\epsilon_B$ is observed for both slip and non-slip boundary conditions. For moderate colloid-solute interaction strength $\epsilon_B$, the difference between slip and non-slip velocities is around 20\% but it becomes smaller with $\epsilon_B$ and eventually becomes vanishingly small.
This observation is not surprising, because the hydrodynamic boundary conditions for a colloid densely coated with much less mobile solute particles should behave like those of a larger colloid with non-slip boundary conditions~\cite{Bocquet94}.

\begin{figure}[h!]
   \begin{center}
      \includegraphics*[width=0.5\textwidth]{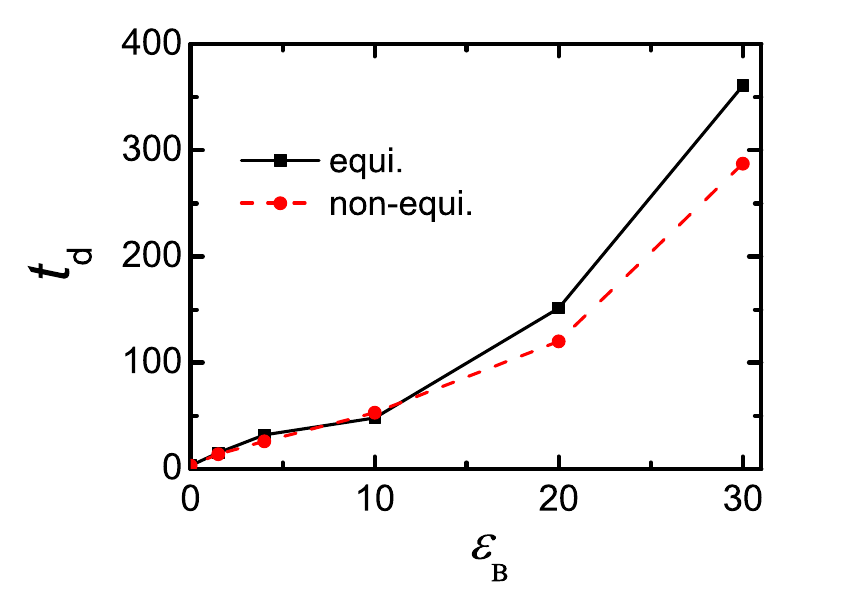}
      \caption{The characteristic time of fluid particle bound by the colloid $t_d$ versus interaction strength between solute and colloid $\epsilon_B$ for equilibrium and non-equilibrium system. In these simulations, slip boundary conditions are imposed.}
      \label{fig:decaytime}
   \end{center}
\end{figure}

To test to what extent attractive forces can bind solute particles to the colloid,  we probe $N_b(t)$,  the number of particles that remain bound to the colloid (i.e. $r_{ic}<r_{cut}$) for at leats  $t$ time-steps. $(N_b(t)/N_b(0))$ decays exponentially with time (see SI).  The rate of decay
is a measure for the characteristic time $t_d$ during which a fluid particle is bound to the colloid. The dependence of $t_d$ on $\epsilon_B$ is shown in Fig.~\ref{fig:decaytime}. As expected, $t_d$ increases with $\epsilon_B$. We also note that $t_d$ decreases with increasing flow rate: the difference woith the equilibrium case is most pronounced for  larger values of $\epsilon_B$.

Figs.~\ref{fig:P0allnew}(a-b) compare the radial distribution of the reduced excess density of the solutes. For both slip and non-slip boundaries, when $\epsilon_B$ is increased from $0.0$ to $1.5$, significant excess of solutes near the colloid is observed. When $\epsilon_B$ is further increased, the solutes close to the colloid are effectively trapped and form a rigid shell at around $r=3.0$ that does not contribute to diffusio-phoresis. The peak is higher and sharper for non-slip boundaries. Fig.~\ref{fig:P0allnew}(c-d) presents the radial profile of fluid velocity $v_x$ with slip and non-slip boundaries. As expected, for $\epsilon_B>1.50$ the magnitude of $v_x(r)$ is decreasing due to the formation of the shell-region, and generally at the same $\epsilon_B$ the magnitude of $v_x$ is smaller for non-slip boundaries, particularly close to the colloid.

\subsection{\label{sec:Res2} Anisotropic interaction}

\begin{figure}[h!]
   \begin{center}
      \includegraphics*[width=0.5\textwidth]{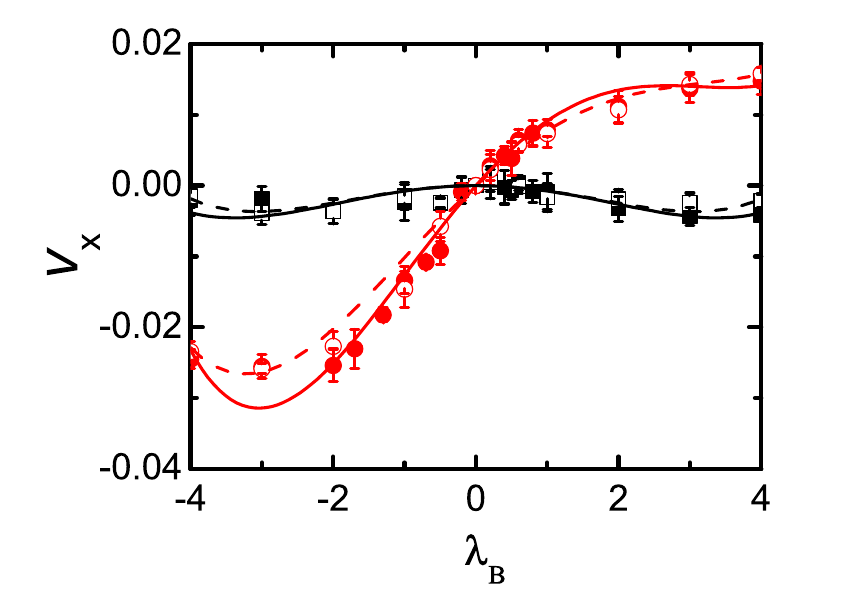}
      \caption{The dependence of $v_x$ on the strength of the anisotropic colloid-solute interaction ($\lambda_B$), for $\epsilon_B=0.0$.  The black curves shows the result for a solute-colloid interaction with $P_1(\cos\theta)$-symmetry. The red curves correspond to the results of an interaction with   $P_2(\cos\theta)$ symmetry. The full symbols and drawn curves correspond to the simulation results for slip boundary conditions, the hollow symbols and dashed curves correspond to the results for non-slip boundary condition. The curves have been obtained as weighted fits (Eqs.~\ref{eqn:fit1} to the simulation data.}
      \label{fig:anglep1p2}
   \end{center}
\end{figure}
To distinguish between different symmetry classes of the anisotropic solute-colloid interactions, we assume that their angular dependence is proportional to Legendre polynomials (\ref{eqn:Aniso}).
We note that the case $l=2$ is unique in the sense that it is the only anisotropic term for which a linear relation between gradient and flow is possible.
For other values of $l$, the flow vanishes in the limit of weak attraction and the phoretic flow velocity should depend at least quadratically on the strength of the colloid-solute attraction. This difference in behavior can be understood by noting that both  $v_x$ and $\nabla\mu$ are polar vectors that transform as irreducible tensors of rank one. In the linear regime (i.e. the regime where $v_x$ depends linearly on $\nabla\mu$), the variation of $v_x$ with $\nabla\mu$ is given by a symmetric matrix. Such a matrix has two parts: the trace, which transforms as a scalar  ($l$=0), and the traceless symmetric part that transforms as second-rank irreducible tensor ($l$=2). For isotropic particles, only the $l=0$ part contributes. However, if the interaction potential is anisotropic, then the $l=2$ part (i.e. the part of the interaction that transforms as $P_2(\cos\theta)$), can also contribute. However, all other angular dependence of the potential ($P_1, P_3, P_4$ etc) do not contribute to linear order. Note, however, that when the interaction becomes stronger, other Legendre components of the potential may contribute to the phoretic flow, because the excess density depends exponentially on the potential. This means that the excess density then no longer has the same symmetry as the anisotropic potential. In that case,  the positive and negative contributions to the excess solute density do no longer cancel, and hence there may still be a flow for, say,  $l=1$ at larger values of $\lambda_B$. We have considered anisotropic interactions of first ($l=1$) and second ($l=2$) order and evaluated the phoretic flow velocity as a function of the magnitude $\lambda_B$.

The results are presented in Fig.~\ref{fig:anglep1p2} for slip (solid curves) and non-slip (dashed curves) boundary conditions.
As expected, at small $\lambda_B$, the flow velocity depends linearly on the interaction strength for $l=2$, but quadratically for $l=1$.
The difference between slip and non-slip boundary conditions is barely significant.
We fit the simulated data with polynomials:
\begin{eqnarray}\label{eqn:fit1}
 v_x = \sum_{i=1}^{i_{max}} A_i \lambda_B ^{i}\;.
\end{eqnarray}
Powers up to fourth order ($i_{max}=4$) need to be considered to obtain a good fit to the data.
Based on the above symmetry arguments, we require that $A_1=A_3\equiv 0$ for $l=1$.
Table~\ref{table_fit} summarizes the fit coefficients for the four cases considered: $l=1,2$ and for slip and non-slip boundaries.
\begin{table}[h!]
\begin{tabular}{|@{ }l@{ }|@{ }c@{ }|@{ }c@{ }|@{ }c@{ }|@{ }c@{ }|}
\hline
 & ${\mathbf A_1}$ & ${\mathbf A_2}$ & ${\mathbf A_3}$ & ${\mathbf A_4}$  \\
\hline
 {\bf l=1}&   &   &   &    \\

 slip & 0 & $-7.94\cdot 10^{-4}$ & 0 & $3.46\cdot 10^{-5}$ \\
 non-slip & 0 & $-7.76\cdot 10^{-4}$ & 0 & $4.13\cdot 10^{-5}$ \\
\hline

 {\bf l=2} &  &  &  &   \\
 slip &
 $1.131\cdot 10^{-2}$ & $-1.84\cdot 10^{-3}$ & $-4.15\cdot 10^{-4}$ & $9.75\cdot 10^{-5}$\\
 non-slip &
 $9.26\cdot 10^{-3}$ & $-1.25\cdot 10^{-3}$ & $-2.77\cdot 10^{-4}$ & $6.40\cdot 10^{-5}$\\
\hline
\end{tabular}
\caption{Fit coefficients $A_i$ in Eqn.~\ref{eqn:fit1} for slip and non-slip boundaries.\label{table_fit}}
\end{table}

\section{\label{sec:Con}Conclusions}
The main findings of the present work are that, although in general the strength of diffusio-phoresis increases with the excess density of adsorbed solute, the diffusio-phoretic motion is suppressed if the binding between colloid and solute is  very strong.
In the limit of strong binding, the adsorbed layer becomes immobile and plays no role in phoresis. However, the phoretic velocity remains finite, even for very strong adsorption, presumably because of the structuring of the fluid around the strongly adsorbed layer.
In addition, we investigated the effect on phoresis of angle-dependent interactions. In analogy with what is found for the electrophoresis of colloids with a thin double layer, we find that the diffusion phoretic effect changes qualitatively as the symmetry of the anisotropic interaction is changed.
These findings demonstrate that the strength of diffusio-phoresis of particles with a patchy interaction (e.g. proteins) depends not just on the strength of the patchy interaction, but on the precise distribution of the patches over the surface of the particle.

Our simulations suggest that although diffusio-phoresis may in principle be used to induce selective transport of bio-macromolecules
in a concentration gradient of selectively binding solutes, the non-monotonic dependence of the transport velocity on the binding strength limits the use of this approach, in the sense that there is no point in making the interaction of the bio-molecules with the solute very strong.  It may then be more useful to selectively adsorb functionalized, charged solutes on the bio-molecules to make them more susceptible to electro-phoresis.
Our simulations suggest that anisotropic interactions may enhance (or decrease) the diffusio-phoretic mobility of an otherwise spherical particle.
However, as only the $l=2$ component of the anisotropic interaction contributes appreciably, the isotropic interaction will dominate the diffusio-phoresis of particles that have a large number of interaction ``patches''.


\section*{Acknowledgements}
This work was partially funded by the National Key Research and Development Program of China (grant 2016YFA0501601), by the EU Horizon 2020 program through 766972-FET-OPEN-NANOPHLOW, by K. C. Wong Educational Foundation, and by the National Natural Science Foundations of China under grants 11602279 and 11874398.

\balance


\bibliography{Wei_et_al_DP}

\providecommand*{\mcitethebibliography}{\thebibliography}
\csname @ifundefined\endcsname{endmcitethebibliography}
{\let\endmcitethebibliography\endthebibliography}{}
\begin{mcitethebibliography}{48}
\providecommand*{\natexlab}[1]{#1}
\providecommand*{\mciteSetBstSublistMode}[1]{}
\providecommand*{\mciteSetBstMaxWidthForm}[2]{}
\providecommand*{\mciteBstWouldAddEndPuncttrue}
  {\def\EndOfBibitem{\unskip.}}
\providecommand*{\mciteBstWouldAddEndPunctfalse}
  {\let\EndOfBibitem\relax}
\providecommand*{\mciteSetBstMidEndSepPunct}[3]{}
\providecommand*{\mciteSetBstSublistLabelBeginEnd}[3]{}
\providecommand*{\EndOfBibitem}{}
\mciteSetBstSublistMode{f}
\mciteSetBstMaxWidthForm{subitem}
{(\emph{\alph{mcitesubitemcount}})}
\mciteSetBstSublistLabelBeginEnd{\mcitemaxwidthsubitemform\space}
{\relax}{\relax}

\bibitem[Eijkel and van~den Berg(2010)]{Eijkel10}
J.~C. Eijkel and A.~van~den Berg, \emph{Chem Soc Rev}, 2010, \textbf{39},
  957--73\relax
\mciteBstWouldAddEndPuncttrue
\mciteSetBstMidEndSepPunct{\mcitedefaultmidpunct}
{\mcitedefaultendpunct}{\mcitedefaultseppunct}\relax
\EndOfBibitem
\bibitem[Anderson and Prieve(2006)]{Anderson06}
J.~L. Anderson and D.~C. Prieve, \emph{Separation and Purification Methods},
  2006, \textbf{13}, 67--103\relax
\mciteBstWouldAddEndPuncttrue
\mciteSetBstMidEndSepPunct{\mcitedefaultmidpunct}
{\mcitedefaultendpunct}{\mcitedefaultseppunct}\relax
\EndOfBibitem
\bibitem[Annunziata \emph{et~al.}(2012)Annunziata, Buzatu, and
  Albright]{Annunziata12}
O.~Annunziata, D.~Buzatu and J.~G. Albright, \emph{J Phys Chem B}, 2012,
  \textbf{116}, 12694--705\relax
\mciteBstWouldAddEndPuncttrue
\mciteSetBstMidEndSepPunct{\mcitedefaultmidpunct}
{\mcitedefaultendpunct}{\mcitedefaultseppunct}\relax
\EndOfBibitem
\bibitem[Ault \emph{et~al.}(2017)Ault, Warren, Shin, and Stone]{Ault17}
J.~T. Ault, P.~B. Warren, S.~Shin and H.~A. Stone, \emph{Soft Matter}, 2017,
  \textbf{13}, 9015--9023\relax
\mciteBstWouldAddEndPuncttrue
\mciteSetBstMidEndSepPunct{\mcitedefaultmidpunct}
{\mcitedefaultendpunct}{\mcitedefaultseppunct}\relax
\EndOfBibitem
\bibitem[Deseigne \emph{et~al.}(2014)Deseigne, Cottin-Bizonne, Stroock,
  Bocquet, and Ybert]{Deseigne14}
J.~Deseigne, C.~Cottin-Bizonne, A.~D. Stroock, L.~Bocquet and C.~Ybert,
  \emph{Soft Matter}, 2014, \textbf{10}, 4795--9\relax
\mciteBstWouldAddEndPuncttrue
\mciteSetBstMidEndSepPunct{\mcitedefaultmidpunct}
{\mcitedefaultendpunct}{\mcitedefaultseppunct}\relax
\EndOfBibitem
\bibitem[Khair(2013)]{Khair13}
A.~S. Khair, \emph{Journal of Fluid Mechanics}, 2013, \textbf{731},
  64--94\relax
\mciteBstWouldAddEndPuncttrue
\mciteSetBstMidEndSepPunct{\mcitedefaultmidpunct}
{\mcitedefaultendpunct}{\mcitedefaultseppunct}\relax
\EndOfBibitem
\bibitem[Palacci \emph{et~al.}(2010)Palacci, Abecassis, Cottin-Bizonne, Ybert,
  and Bocquet]{Palacci10}
J.~Palacci, B.~Abecassis, C.~Cottin-Bizonne, C.~Ybert and L.~Bocquet,
  \emph{Phys Rev Lett}, 2010, \textbf{104}, 138302\relax
\mciteBstWouldAddEndPuncttrue
\mciteSetBstMidEndSepPunct{\mcitedefaultmidpunct}
{\mcitedefaultendpunct}{\mcitedefaultseppunct}\relax
\EndOfBibitem
\bibitem[Paustian \emph{et~al.}(2015)Paustian, Angulo, Nery-Azevedo, Shi,
  Abdel-Fattah, and Squires]{Paustian15}
J.~S. Paustian, C.~D. Angulo, R.~Nery-Azevedo, N.~Shi, A.~I. Abdel-Fattah and
  T.~M. Squires, \emph{Langmuir}, 2015, \textbf{31}, 4402--10\relax
\mciteBstWouldAddEndPuncttrue
\mciteSetBstMidEndSepPunct{\mcitedefaultmidpunct}
{\mcitedefaultendpunct}{\mcitedefaultseppunct}\relax
\EndOfBibitem
\bibitem[Prieve \emph{et~al.}(2019)Prieve, Malone, Khair, Stout, and
  Kanj]{Prieve18}
D.~C. Prieve, S.~M. Malone, A.~S. Khair, R.~F. Stout and M.~Y. Kanj, \emph{Proc
  Natl Acad Sci U S A}, 2019, \textbf{116}, 18257--18262\relax
\mciteBstWouldAddEndPuncttrue
\mciteSetBstMidEndSepPunct{\mcitedefaultmidpunct}
{\mcitedefaultendpunct}{\mcitedefaultseppunct}\relax
\EndOfBibitem
\bibitem[Sear and Warren(2017)]{Sear17}
R.~P. Sear and P.~B. Warren, \emph{Phys Rev E}, 2017, \textbf{96}, 062602\relax
\mciteBstWouldAddEndPuncttrue
\mciteSetBstMidEndSepPunct{\mcitedefaultmidpunct}
{\mcitedefaultendpunct}{\mcitedefaultseppunct}\relax
\EndOfBibitem
\bibitem[Shi \emph{et~al.}(2016)Shi, Nery-Azevedo, Abdel-Fattah, and
  Squires]{Shi16}
N.~Shi, R.~Nery-Azevedo, A.~I. Abdel-Fattah and T.~M. Squires, \emph{Phys Rev
  Lett}, 2016, \textbf{117}, 258001\relax
\mciteBstWouldAddEndPuncttrue
\mciteSetBstMidEndSepPunct{\mcitedefaultmidpunct}
{\mcitedefaultendpunct}{\mcitedefaultseppunct}\relax
\EndOfBibitem
\bibitem[Shin \emph{et~al.}(2017)Shin, Ault, Warren, and Stone]{Shin17}
S.~Shin, J.~T. Ault, P.~B. Warren and H.~A. Stone, \emph{Physical Review X},
  2017, \textbf{7}, year\relax
\mciteBstWouldAddEndPuncttrue
\mciteSetBstMidEndSepPunct{\mcitedefaultmidpunct}
{\mcitedefaultendpunct}{\mcitedefaultseppunct}\relax
\EndOfBibitem
\bibitem[Velegol \emph{et~al.}(2016)Velegol, Garg, Guha, Kar, and
  Kumar]{Velegol16}
D.~Velegol, A.~Garg, R.~Guha, A.~Kar and M.~Kumar, \emph{Soft Matter}, 2016,
  \textbf{12}, 4686--703\relax
\mciteBstWouldAddEndPuncttrue
\mciteSetBstMidEndSepPunct{\mcitedefaultmidpunct}
{\mcitedefaultendpunct}{\mcitedefaultseppunct}\relax
\EndOfBibitem
\bibitem[Yang \emph{et~al.}(2014)Yang, Wysocki, and Ripoll]{Yang14}
M.~Yang, A.~Wysocki and M.~Ripoll, \emph{Soft Matter}, 2014, \textbf{10},
  6208--6218\relax
\mciteBstWouldAddEndPuncttrue
\mciteSetBstMidEndSepPunct{\mcitedefaultmidpunct}
{\mcitedefaultendpunct}{\mcitedefaultseppunct}\relax
\EndOfBibitem
\bibitem[Wurger(2010)]{Wurger10}
A.~Wurger, \emph{Reports on Progress in Physics}, 2010, \textbf{73},
  126601\relax
\mciteBstWouldAddEndPuncttrue
\mciteSetBstMidEndSepPunct{\mcitedefaultmidpunct}
{\mcitedefaultendpunct}{\mcitedefaultseppunct}\relax
\EndOfBibitem
\bibitem[Braibanti \emph{et~al.}(2008)Braibanti, Vigolo, and
  Piazza]{Braibanti08}
M.~Braibanti, D.~Vigolo and R.~Piazza, \emph{Phys Rev Lett}, 2008,
  \textbf{100}, 108303\relax
\mciteBstWouldAddEndPuncttrue
\mciteSetBstMidEndSepPunct{\mcitedefaultmidpunct}
{\mcitedefaultendpunct}{\mcitedefaultseppunct}\relax
\EndOfBibitem
\bibitem[Burelbach \emph{et~al.}(2017)Burelbach, Frenkel, Pagonabarraga, and
  Eiser]{Burelbach17}
J.~Burelbach, D.~Frenkel, I.~Pagonabarraga and E.~Eiser, \emph{The European
  Physical Journal E}, 2017, \textbf{41}, 7\relax
\mciteBstWouldAddEndPuncttrue
\mciteSetBstMidEndSepPunct{\mcitedefaultmidpunct}
{\mcitedefaultendpunct}{\mcitedefaultseppunct}\relax
\EndOfBibitem
\bibitem[Burelbach \emph{et~al.}(2017)Burelbach, Zupkauskas, Lamboll, Lan, and
  Eiser]{Burelbach17-2}
J.~Burelbach, M.~Zupkauskas, R.~Lamboll, Y.~Lan and E.~Eiser, \emph{J Chem
  Phys}, 2017, \textbf{147}, 094906\relax
\mciteBstWouldAddEndPuncttrue
\mciteSetBstMidEndSepPunct{\mcitedefaultmidpunct}
{\mcitedefaultendpunct}{\mcitedefaultseppunct}\relax
\EndOfBibitem
\bibitem[Fu \emph{et~al.}(2017)Fu, Merabia, and Joly]{Fu17}
L.~Fu, S.~Merabia and L.~Joly, \emph{Phys Rev Lett}, 2017, \textbf{119},
  214501\relax
\mciteBstWouldAddEndPuncttrue
\mciteSetBstMidEndSepPunct{\mcitedefaultmidpunct}
{\mcitedefaultendpunct}{\mcitedefaultseppunct}\relax
\EndOfBibitem
\bibitem[Maeda \emph{et~al.}(2011)Maeda, Buguin, and Libchaber]{Maeda11}
Y.~T. Maeda, A.~Buguin and A.~Libchaber, \emph{Phys Rev Lett}, 2011,
  \textbf{107}, 038301\relax
\mciteBstWouldAddEndPuncttrue
\mciteSetBstMidEndSepPunct{\mcitedefaultmidpunct}
{\mcitedefaultendpunct}{\mcitedefaultseppunct}\relax
\EndOfBibitem
\bibitem[Piazza(2008)]{Piazza08}
R.~Piazza, \emph{Soft Matter}, 2008, \textbf{4}, 1740\relax
\mciteBstWouldAddEndPuncttrue
\mciteSetBstMidEndSepPunct{\mcitedefaultmidpunct}
{\mcitedefaultendpunct}{\mcitedefaultseppunct}\relax
\EndOfBibitem
\bibitem[Piazza and Parola(2008)]{Piazza08-2}
R.~Piazza and A.~Parola, \emph{Journal of Physics: Condensed Matter}, 2008,
  \textbf{20}, 153102\relax
\mciteBstWouldAddEndPuncttrue
\mciteSetBstMidEndSepPunct{\mcitedefaultmidpunct}
{\mcitedefaultendpunct}{\mcitedefaultseppunct}\relax
\EndOfBibitem
\bibitem[Tsuji \emph{et~al.}(2017)Tsuji, Kozai, Ishino, and Kawano]{Tsuji17}
T.~Tsuji, K.~Kozai, H.~Ishino and S.~Kawano, \emph{Micro and Nano Letters},
  2017, \textbf{12}, 520--525\relax
\mciteBstWouldAddEndPuncttrue
\mciteSetBstMidEndSepPunct{\mcitedefaultmidpunct}
{\mcitedefaultendpunct}{\mcitedefaultseppunct}\relax
\EndOfBibitem
\bibitem[Vigolo \emph{et~al.}(2010)Vigolo, Rusconi, Stone, and
  Piazza]{Vigolo10}
D.~Vigolo, R.~Rusconi, H.~A. Stone and R.~Piazza, \emph{Soft Matter}, 2010,
  \textbf{6}, 3489\relax
\mciteBstWouldAddEndPuncttrue
\mciteSetBstMidEndSepPunct{\mcitedefaultmidpunct}
{\mcitedefaultendpunct}{\mcitedefaultseppunct}\relax
\EndOfBibitem
\bibitem[Vigolo \emph{et~al.}(2017)Vigolo, Zhao, Handschin, Cao, deMello, and
  Mezzenga]{Vigolo17}
D.~Vigolo, J.~Zhao, S.~Handschin, X.~Cao, A.~J. deMello and R.~Mezzenga,
  \emph{Sci Rep}, 2017, \textbf{7}, 1211\relax
\mciteBstWouldAddEndPuncttrue
\mciteSetBstMidEndSepPunct{\mcitedefaultmidpunct}
{\mcitedefaultendpunct}{\mcitedefaultseppunct}\relax
\EndOfBibitem
\bibitem[Wienken \emph{et~al.}(2010)Wienken, Baaske, Rothbauer, Braun, and
  Duhr]{Wienken10}
C.~J. Wienken, P.~Baaske, U.~Rothbauer, D.~Braun and S.~Duhr, \emph{Nat
  Commun}, 2010, \textbf{1}, 100\relax
\mciteBstWouldAddEndPuncttrue
\mciteSetBstMidEndSepPunct{\mcitedefaultmidpunct}
{\mcitedefaultendpunct}{\mcitedefaultseppunct}\relax
\EndOfBibitem
\bibitem[Wolff \emph{et~al.}(2016)Wolff, Mittag, Herling, Genst, Dobson,
  Knowles, Braun, and Buell]{Wolff16}
M.~Wolff, J.~J. Mittag, T.~W. Herling, E.~D. Genst, C.~M. Dobson, T.~P.
  Knowles, D.~Braun and A.~K. Buell, \emph{Sci Rep}, 2016, \textbf{6},
  22829\relax
\mciteBstWouldAddEndPuncttrue
\mciteSetBstMidEndSepPunct{\mcitedefaultmidpunct}
{\mcitedefaultendpunct}{\mcitedefaultseppunct}\relax
\EndOfBibitem
\bibitem[Molotilin \emph{et~al.}(2016)Molotilin, Lobaskin, and
  Vinogradova]{Molotilin16}
T.~Y. Molotilin, V.~Lobaskin and O.~I. Vinogradova, \emph{J Chem Phys}, 2016,
  \textbf{145}, 244704\relax
\mciteBstWouldAddEndPuncttrue
\mciteSetBstMidEndSepPunct{\mcitedefaultmidpunct}
{\mcitedefaultendpunct}{\mcitedefaultseppunct}\relax
\EndOfBibitem
\bibitem[Semenov \emph{et~al.}(2013)Semenov, Raafatnia, Sega, Lobaskin, Holm,
  and Kremer]{Semenov13}
I.~Semenov, S.~Raafatnia, M.~Sega, V.~Lobaskin, C.~Holm and F.~Kremer,
  \emph{Phys Rev E Stat Nonlin Soft Matter Phys}, 2013, \textbf{87},
  022302\relax
\mciteBstWouldAddEndPuncttrue
\mciteSetBstMidEndSepPunct{\mcitedefaultmidpunct}
{\mcitedefaultendpunct}{\mcitedefaultseppunct}\relax
\EndOfBibitem
\bibitem[Stout and Khair(2017)]{Stout17}
R.~F. Stout and A.~S. Khair, \emph{Phys. Rev. Fluids}, 2017, \textbf{2},
  014201\relax
\mciteBstWouldAddEndPuncttrue
\mciteSetBstMidEndSepPunct{\mcitedefaultmidpunct}
{\mcitedefaultendpunct}{\mcitedefaultseppunct}\relax
\EndOfBibitem
\bibitem[Tanaka and Grosberg(2002)]{Tanaka02}
M.~Tanaka and A.~Y. Grosberg, \emph{Eur Phys J E Soft Matter}, 2002,
  \textbf{7}, 371--9\relax
\mciteBstWouldAddEndPuncttrue
\mciteSetBstMidEndSepPunct{\mcitedefaultmidpunct}
{\mcitedefaultendpunct}{\mcitedefaultseppunct}\relax
\EndOfBibitem
\bibitem[Tanaka(2003)]{Tanaka03}
M.~Tanaka, \emph{Phys Rev E Stat Nonlin Soft Matter Phys}, 2003, \textbf{68},
  061501\relax
\mciteBstWouldAddEndPuncttrue
\mciteSetBstMidEndSepPunct{\mcitedefaultmidpunct}
{\mcitedefaultendpunct}{\mcitedefaultseppunct}\relax
\EndOfBibitem
\bibitem[Anderson(1981)]{Anderson81}
J.~L. Anderson, \emph{Journal of Colloid and Interface Science}, 1981,
  \textbf{82}, 248--250\relax
\mciteBstWouldAddEndPuncttrue
\mciteSetBstMidEndSepPunct{\mcitedefaultmidpunct}
{\mcitedefaultendpunct}{\mcitedefaultseppunct}\relax
\EndOfBibitem
\bibitem[Yang \emph{et~al.}(2017)Yang, Liu, Ye, and Chen]{Yang17}
M.~Yang, R.~Liu, F.~Ye and K.~Chen, \emph{Soft Matter}, 2017, \textbf{13},
  647--657\relax
\mciteBstWouldAddEndPuncttrue
\mciteSetBstMidEndSepPunct{\mcitedefaultmidpunct}
{\mcitedefaultendpunct}{\mcitedefaultseppunct}\relax
\EndOfBibitem
\bibitem[Chen \emph{et~al.}(2017)Chen, Xu, and Ren]{Chen17}
T.~Chen, C.~Xu and Z.~Ren, \emph{Journal of Industrial Management
  Optimization}, 2017, \textbf{13}, 1--19\relax
\mciteBstWouldAddEndPuncttrue
\mciteSetBstMidEndSepPunct{\mcitedefaultmidpunct}
{\mcitedefaultendpunct}{\mcitedefaultseppunct}\relax
\EndOfBibitem
\bibitem[Huang \emph{et~al.}(2016)Huang, Schofield, and Kapral]{Huang16}
M.-J. Huang, J.~Schofield and R.~Kapral, \emph{Soft Matter}, 2016, \textbf{12},
  5581--5589\relax
\mciteBstWouldAddEndPuncttrue
\mciteSetBstMidEndSepPunct{\mcitedefaultmidpunct}
{\mcitedefaultendpunct}{\mcitedefaultseppunct}\relax
\EndOfBibitem
\bibitem[Liu \emph{et~al.}(2018)Liu, Ganti, and Frenkel]{Liu18}
Y.~Liu, R.~Ganti and D.~Frenkel, \emph{Journal of Physics: Condensed Matter},
  2018, \textbf{30}, 205002\relax
\mciteBstWouldAddEndPuncttrue
\mciteSetBstMidEndSepPunct{\mcitedefaultmidpunct}
{\mcitedefaultendpunct}{\mcitedefaultseppunct}\relax
\EndOfBibitem
\bibitem[Sharifi-Mood \emph{et~al.}(2013)Sharifi-Mood, Koplik, and
  Maldarelli]{Sharifi-Mood13}
N.~Sharifi-Mood, J.~Koplik and C.~Maldarelli, \emph{Phys Rev Lett}, 2013,
  \textbf{111}, 184501\relax
\mciteBstWouldAddEndPuncttrue
\mciteSetBstMidEndSepPunct{\mcitedefaultmidpunct}
{\mcitedefaultendpunct}{\mcitedefaultseppunct}\relax
\EndOfBibitem
\bibitem[Abecassis \emph{et~al.}(2008)Abecassis, Cottin-Bizonne, Ybert, Ajdari,
  and Bocquet]{Abecassis08}
B.~Abecassis, C.~Cottin-Bizonne, C.~Ybert, A.~Ajdari and L.~Bocquet, \emph{Nat
  Mater}, 2008, \textbf{7}, 785--9\relax
\mciteBstWouldAddEndPuncttrue
\mciteSetBstMidEndSepPunct{\mcitedefaultmidpunct}
{\mcitedefaultendpunct}{\mcitedefaultseppunct}\relax
\EndOfBibitem
\bibitem[Banerjee \emph{et~al.}(2016)Banerjee, Williams, Azevedo, Helgeson, and
  Squires]{Banerjee16}
A.~Banerjee, I.~Williams, R.~N. Azevedo, M.~E. Helgeson and T.~M. Squires,
  \emph{Proceedings of the National Academy of Sciences}, 2016, \textbf{113},
  8612--8617\relax
\mciteBstWouldAddEndPuncttrue
\mciteSetBstMidEndSepPunct{\mcitedefaultmidpunct}
{\mcitedefaultendpunct}{\mcitedefaultseppunct}\relax
\EndOfBibitem
\bibitem[Saar \emph{et~al.}(2017)Saar, Zhang, Muller, Kumar, Devenish, Lynn,
  Lapinska, Yang, Linse, and Knowles]{Saar17}
K.~L. Saar, Y.~Zhang, T.~Muller, C.~P. Kumar, S.~Devenish, A.~Lynn,
  U.~Lapinska, X.~Yang, S.~Linse and T.~P.~J. Knowles, \emph{Lab Chip}, 2017,
  \textbf{18}, 162--170\relax
\mciteBstWouldAddEndPuncttrue
\mciteSetBstMidEndSepPunct{\mcitedefaultmidpunct}
{\mcitedefaultendpunct}{\mcitedefaultseppunct}\relax
\EndOfBibitem
\bibitem[Yang and Ripoll(2016)]{Yang16}
M.~Yang and M.~Ripoll, \emph{Soft Matter}, 2016, \textbf{12}, 8564--8573\relax
\mciteBstWouldAddEndPuncttrue
\mciteSetBstMidEndSepPunct{\mcitedefaultmidpunct}
{\mcitedefaultendpunct}{\mcitedefaultseppunct}\relax
\EndOfBibitem
\bibitem[Liu \emph{et~al.}(2017)Liu, Ganti, Burton, Zhang, Wang, and
  Frenkel]{Liu17}
Y.~Liu, R.~Ganti, H.~G. Burton, X.~Zhang, W.~Wang and D.~Frenkel,
  \emph{Physical Review Letters}, 2017, \textbf{119}, 224502\relax
\mciteBstWouldAddEndPuncttrue
\mciteSetBstMidEndSepPunct{\mcitedefaultmidpunct}
{\mcitedefaultendpunct}{\mcitedefaultseppunct}\relax
\EndOfBibitem
\bibitem[Yoshida \emph{et~al.}(2017)Yoshida, Marbach, and Bocquet]{Yoshida17}
H.~Yoshida, S.~Marbach and L.~Bocquet, \emph{The Journal of Chemical Physics},
  2017, \textbf{146}, 194702\relax
\mciteBstWouldAddEndPuncttrue
\mciteSetBstMidEndSepPunct{\mcitedefaultmidpunct}
{\mcitedefaultendpunct}{\mcitedefaultseppunct}\relax
\EndOfBibitem
\bibitem[Ganti \emph{et~al.}(2017)Ganti, Liu, and Frenkel]{Ganti17}
R.~Ganti, Y.~Liu and D.~Frenkel, \emph{Phys Rev Lett}, 2017, \textbf{119},
  038002\relax
\mciteBstWouldAddEndPuncttrue
\mciteSetBstMidEndSepPunct{\mcitedefaultmidpunct}
{\mcitedefaultendpunct}{\mcitedefaultseppunct}\relax
\EndOfBibitem
\bibitem[Wang \emph{et~al.}(2019)Wang, Ram\'irez-Hinestrosa, Dobnikar, and
  Frenkel]{Wang2019}
X.~Wang, S.~Ram\'irez-Hinestrosa, J.~Dobnikar and D.~Frenkel, \emph{The
  Lennard-Jones potential: when (not) to use it}, 2019\relax
\mciteBstWouldAddEndPuncttrue
\mciteSetBstMidEndSepPunct{\mcitedefaultmidpunct}
{\mcitedefaultendpunct}{\mcitedefaultseppunct}\relax
\EndOfBibitem
\bibitem[Ram\'irez-Hinestrosa \emph{et~al.}(2019)Ram\'irez-Hinestrosa, Yoshida,
  Bocquet, and Frenkel]{Ramirez2019}
S.~Ram\'irez-Hinestrosa, H.~Yoshida, L.~Bocquet and D.~Frenkel, \emph{Numerical
  analysis of polymer diffusiophoresis by means of the molecular dynamics},
  2019\relax
\mciteBstWouldAddEndPuncttrue
\mciteSetBstMidEndSepPunct{\mcitedefaultmidpunct}
{\mcitedefaultendpunct}{\mcitedefaultseppunct}\relax
\EndOfBibitem
\bibitem[Bocquet and Barrat({1994})]{Bocquet94}
L.~Bocquet and J.~Barrat, \emph{{Phys. Rev. E}}, {1994}, \textbf{{49}},
  {3079--3092}\relax
\mciteBstWouldAddEndPuncttrue
\mciteSetBstMidEndSepPunct{\mcitedefaultmidpunct}
{\mcitedefaultendpunct}{\mcitedefaultseppunct}\relax
\EndOfBibitem
\end{mcitethebibliography}
\bibliographystyle{rsc}
\end{document}